\documentclass[12pt,a4paper]{article}

\newcommand{\bq}{\mbox{\boldmath$q$}}
\newcommand{\bk}{\mbox{\boldmath$k$}}
\newcommand{\bp}{\mbox{\boldmath$p$}}
\newcommand{\be}{\mbox{\boldmath$e$}}
\newcommand{\br}{\mbox{\boldmath$r$}}

\begin{document}

\title{Triple coalescence singularity in a dynamical
atomic process}

\author{E. G. Drukarev$^{1,2}, $ A. I. Mikhailov$^{1,2}$,
I. A. Mikhailov$^{1,3}$, and  W. Scheid$^2$\\[0.2ex]
\em $^1$ Petersburg Nuclear Physics Institute,\\
\em Gatchina, St. Petersburg 188300, Russia\\
\em $^2$ Justus-Liebig-Universit\"at Giessen, Giessen 35392, Germany \\
\em  $^3$ University of Central Florida, Orlando, Florida 32816, USA}

\date{}
\maketitle

\begin{abstract}
We show that the high energy limit for amplitude of the double
electron capture to the bound state of the Coulomb field of a
nucleus with emission of a single photon is determined by behavior
of the wave function in the vicinity of the singular triple
coalescence point. \\
PCAC: 31.25.Eb; 32.80.-t
\end{abstract}

It is well known that the solution $\Psi(\br_1,\br_2)$ of the
Schr\"odinger equation for a two-electron system in the Coulomb
field of infinitely heavy point nucleus is singular at the triple
coalescence point $\br_1=\br_2=0$ \cite{1,2,3,4}. The wave
function can be presented as a generalized power series containing
logarithmic terms. The behavior of $\Psi(\br_1,\br_2)$ near this
point is not very important in calculations of the binding energy
since the corresponding phase volume is small. On the contrary it
becomes crucial for calculation of the local energy
$E(\br_1,\br_2)=H\Psi(\br_1,\br_2)\big/\Psi(\br_1,\br_2)$
\cite{4}. However the triple coalescence point did not manifest
itself in a dynamical process until now.

 In this letter we present an observable effect for the
first time in which the interesting behavior of the two electron
wave function $\Psi(\br_1,\br_2)$ near the point $\br_1=\br_2=0$
(as well as that near the double coalescence points $\br_1=0$ and
$\br_2=0$) plays an important role. For that we consider the
double electron capture followed by the emission of a single
photon in the high energy limit. Since the first attempts to
detect this process in collisions of a light atom with a heavy
nucleus \cite{5} a number of experimental \cite{6,7} and
theoretical \cite{8,9,10} papers were devoted to this reaction.

Neglecting the internal motion of the electrons in the light atom
we consider the capture of two continuum electrons with equal
linear momenta $\bp_1=\bp_2=\bp$ \cite{8}. The process is
characterized by the kinetic energy per nucleon $E_N\,$(MeV/u).
The corresponding electron kinetic energies are
$\varepsilon=E_Nm/m_N$  where $m/m_N$ is the ratio of the electron
mass to the nucleon mass. We consider the case corresponding to
nonrelativistic continuum electrons $\varepsilon\ll m$, i.e.
$E_N\ll m_N$ (in the system of units with $\hbar=c=1$), thus we
put $p=(2m\varepsilon)^{1/2}$. We assume the charge $Z$ of the
heavy nucleus  to be much larger than that of the light atom
$Z_1$, i.e. $Z\gg Z_1$. On the other hand we assume $Z$ to be
small enough for the description of the bound state by the
Schr\"odinger equation, i.e. $(\alpha Z)^2\ll1$, with
$\alpha=1/137$ the fine structure constant.

The energy of the emitted photon is $\omega=2\varepsilon+I$ with
$I>0$ standing for the binding energy of the two electrons in the
ground state of the heavy nucleus (we consider this very case).
 The momentum $\bq=2\bp-\bk$ with $\bk$ standing for
the photon momentum is transferred to the nucleus. We consider the
high-energy limit of the chosen process
\begin{equation}
\varepsilon\ \gg\ I_1\ ,
\label{1}
\end{equation}
with $I_1$ standing for the single-electron binding energy. Since
$Z\gg1$ we can put $I_1=\eta^2/2m$ with $\eta=m\alpha Z$ being the
characteristic momentum of the bound $1s$ state. Thus condition
(\ref{1}) is equivalent to $\xi^2\ll1$ for
\begin{equation}
\xi\ =\ \frac{m\alpha Z}p\,.
\label{2}
\end{equation}

The parameter $\xi$ describes also the
interaction between the incoming electrons and the nucleus. For
$\xi^2\ll1$ this interaction can be treated
perturbatively. Thus the  electronic wave function of
the initial state can be obtained by  an
iteration of the Lippmann--Schwinger equation
\begin{equation}
\Phi\ =\ \Phi_0+GV\Phi\,,
\label{3}
\end{equation}
with $V=V_{eN}+V_{ee}$, while $V_{eN}$ and $V_{ee}$ stand for
 the  interactions between an electron and the nucleus
and between the electrons correspondingly, $G$ is the Green
function of two free noninteracting electrons, while $\Phi_0$ is
the product of plane waves.

Due to conditions (\ref{1}) the electrons transfer  the
large momentum $q\gg\eta$ to the nucleus. This can take place in
 the initial or  final states. The two mechanisms
provide contributions of the same order of magnitude to the
amplitude
\begin{equation}
F\ =\ \langle\Psi|\gamma|\Phi\rangle\,,
\label{4}
\end{equation}
where $\gamma$ is the operator of the interaction between
the electrons and the photon, for which we assume the gradient
form. Each act of exchange by the large momentum
$q\gg\eta$ between the continuum electrons and the
nucleus provides a small factor $\eta^4/q^4$ \cite{11}
since each of the functions $G$ and $V$ on the right-hand side of
Eq.~(\ref{3}) drops as $1/q^2$.
(For the Coulomb case this can be shown
explicitly since the Fourier transform of the wave function drops
as $q^{-4}$). Thus to obtain the lowest order of expansion of the
amplitude (\ref{4}) in powers of $\xi$ we must include
the two lowest terms of iteration of Eq.~(\ref{3}) putting
\begin{equation}
\Phi=\Phi_0+\Phi_1\,; \quad \Phi_1=\Phi_{1N}+\Phi_{1e}\,;\quad
\Phi_{1N}=GV_{eN}\Phi_0\,;\quad \Phi_{1e}=GV_{ee}\Phi_0\,.
\label{5}
\end{equation}
Using Eq. (\ref{5}) we can write
\begin{eqnarray}
&& F=F_0+F_1\,; \quad F_1=F_{1N}+F_{1e}\,; \quad
F_0=\langle\Psi|\gamma|\Phi_0\rangle\,;
\nonumber\\
&& F_{1N}=\langle\Psi|\gamma|\Phi_{1N}\rangle\,; \quad
F_{1e}=\langle\Psi|\gamma|\Phi_{1e}\rangle\,,
\label{6}
\end{eqnarray}
where we expect $F_0$ and $F_1$ to provide contributions of the same
order.

 Let us start with the calculation of the amplitude
$F_0$. Denote $\varphi_0(i,\bp_i)=e^{i(\bp_i\br_i)}$, then
$\Phi_0=\varphi_0(1,\bp_1)\varphi_0(2,\bp_2)$ (recall that in our
case $\bp_1=\bp_2$). Below we assume that the electron which emits
the photon is labelled by ``1". Thus
\begin{equation}
F_0=2\int d^3 r_1d^3 r_2\Psi(\br_1,\br_2)\gamma_1\varphi_0(1,\bp)
\varphi_0(2,p)
\label{7}
\end{equation}
with the operator $\gamma_1$ acting on $\br_1$. Since
$p\gg\eta$, the integral (\ref{7}) over $r_2$ is
saturated by $r_2\sim\frac1p\ll\frac1\eta$, while $\frac1\eta$ is
the characteristic scale of the bound state wave function $\Psi$.
Keeping $r_1$ to be finite we can expand the function
$\Psi(\br_1,\br_2)$ near the point $r_2=0$. Limiting ourselves to
the linear terms of the expansion we can write
\begin{equation}
\Psi(\br_1,\br_2)=\left(1+r_2\frac\partial{\partial r_2}
-\frac{(\br_1\br_2)}{r_1}\frac\partial{\partial\rho}\right)
\tilde\Psi(r_1,r_2,\rho)\bigg|_{r_2=0}
\label{8}
\end{equation}
with $\tilde\Psi(r_1,r_2,\rho)=\Psi(\br_1,\br_2)$,
$\rho=|\br_1-\br_2|$. Following \cite{11} we can present
\[
\tilde\Psi(r_1,0,r_1)=\lim_{\lambda\to0}\tilde\Psi(r_1,0,r_1)
e^{-\lambda r_2}
\]
and obtain for the corresponding  contribution
 $F_{0lin}$ to $F_0$ (lower  index $``lin"$
corresponds to the terms linear in $r_2$ in (\ref{8}))
\begin{equation}
\label{9}
F_{0lin}=-\frac{16\,\pi(\be\bp)}{p^4} J_0\,; \quad J_0=\int
d^3r_1\tilde\Psi^\prime_{r_2}(r_1,0,r_1)e^{i(\bp\br_1)},
\end{equation}
with $\be$ the vector of the photon polarization,
$\tilde\Psi^\prime_{r_2}$ denotes the partial derivative of the
function $\tilde\Psi$ with respect to $r_2$ at $r_2=0$.

The contribution $F_{1N}$ to the amplitude is composed by the
lowest order Coulomb corrections to the plane waves, describing
continuum wave functions in the amplitude $F_0$. Due to the
operator $\gamma$  the correction to the wave function
of the electron ``1" contains  a small factor of the
order $\xi$. Thus the amplitude $F_{1N}$ can be presented as
\begin{equation}
F_{1N}=2\int\frac{d^3p'}{(2\pi)^3}\frac{\langle\Psi|\gamma_1|\varphi_0
(1,\bp)\varphi_0(2,\bp')\rangle\langle\varphi_0(2,\bp')|V_{eN}(2)|
\varphi_0(2,\bp)\rangle}{\varepsilon-\,p'\,^2/2m}
\end{equation}
with $V_{eN}(2)$ standing for the interaction between
the electron ``2" and the nucleus. Straightforward calculation
provides
\begin{equation}
F_{1N}=-\frac{16\pi\eta(\be\cdot\bp)}{p^4}\,J_1\,; \quad J_1=\int
d^3 r_1\tilde\Psi(r_1,0,r_1)e^{i(\bp\cdot\br_1)}.
\label{11}
\end{equation}
Thus
\begin{equation}
F_{0lin}+F_{1N}=-\frac{16\pi(\be\cdot\bp)}{p^4}(J_0+\eta J_1)=0\,.
\label{12}
\end{equation}
 Now, the last equation is due to the Kato cusp condition
$\tilde\Psi_{r_2}'(r_1,0,r_1)=-\eta\tilde\Psi(r_1,0,r_1)$
\cite{12} which can be viewed as the result of cancellation of the
terms $1/r_2$ in the Schr\"odinger equation.

The value of $J_1$ can be obtained by integrating  (11)
three times  in parts. This provides
$J_1=-\frac{8\pi}{p^2}\varphi'_{r_1}(r_1=0)$ with
$\varphi(r_1)=\tilde\Psi(r_1,0,r_1)$. Using the expansion
\begin{equation}
\tilde\Psi(r_1,r_2,\rho)=\left(1-\eta r_1-\eta
r_2+\frac{m\alpha}2\,\rho+\ldots\right)\tilde\Psi(0,0,0)
\label{13}
\end{equation}
at $r_{1}, r_{2}\to0$ (the dots stand for nonlinear
terms) \cite{2,3,4} we find $J_1=8\pi\alpha ZN^2/p^4$, with
$N^2=\tilde\Psi(0,0,0)$  as the value of the wave
function at the origin. Thus the $Z$ dependence of the
amplitudes $F_{0lin}$ and $F_{1N}$ is expressed by the factor
$Z^2N^2\sim Z^5$. Both amplitudes $F_{0lin}$ and $F_{1N}$ describe the
contributions in which each of the electrons transfers
momentum of the order $p\gg\eta$ to the nucleus, providing a
factor $\sim\alpha Z$ to the total amplitude. According
to (12) such contributions cancel.

However there is another contribution to the amplitude $F_{eN}$ in
which one of the captured electrons looses its large momentum $p$
by transferring it to the second electron. The latter transfers
the momentum $2\bp$ to the nucleus. Such contribution
has the same energy dependence as the amplitude, depending on $Z$
as $Z^4$, since the electron interaction with the nucleus,
proportional to $\alpha Z$ is replaced by the
electron--electron interaction which is proportional to $\alpha$.
This contribution is described by the nonlinear terms
of expansion (\ref{13}) for the wave function denoted by dots on
the right-hand side. However due to the singularity at the point
$r_1=r_2=\rho=0$  the expansion (\ref{13}) beyond the
linear terms is not the Taylor series and the coefficients cannot
be presented in terms of partial derivatives. The leading
nonlinear terms are obtained in \cite{2,3,4}.

 The contribution $F_{1e}$ can be calculated in similar
way. It also behaves as $Z^4$, e.g.
\begin{equation}
F_{1e}\ =\ -\frac{16\pi^2\alpha\eta N^2(\be\cdot\bp)}{p^7}\,.
\label{14}
\end{equation}
Finally, the high energy limit of the amplitude is
\begin{equation}
F=2\left(\int d^3r_1d^3r_2\Psi(\br_1,\br_2)\gamma_1e^{i(\bp_1\br_1)
+i(\bp_2\br_2)} -16Z\Lambda-2\Lambda\right),
\label{15}
\end{equation}
with $\bp_1=\bp_2=\bp$,
$\Lambda=4\pi^2\alpha\eta N^2(\be\cdot\bp)/p^8$. The three terms in
brackets correspond to the contributions $F_0$, $F_{1N}$ and $F_{1e}$.
Since the integral is determined by $r_1\sim r_2\sim p^{-1}\to0$, one
should use generalized power series found in \cite{2,3,4} for
calculations.

Equation (\ref{15}) is true for $\omega\ll\eta$. At
larger values of photon energies there are some additional terms
corresponding to the two-step mechanism \cite{13} in which the
process can be viewed as the $eN$ scattering followed by quasifree
emission of the photon. The effects of internal motion of the
target electrons \cite{10} (in experiments the light atom is the
target) provide the contributions of the order $m\alpha Z_1/p$
being thus beyond the high energy limit. These effects
can be incorporated into Eq.~(\ref{15}).

In conclusion we recognize that the amplitude $F$ in Eq.\,(15)
depends sensitively on the two electron wave function at the
points $\br_1=0$, $\br_2=0$ and $\br_1=\br_2=0$. So the
measurement of the cross sections for double electron capture with
emission of a single photon in the high energy limit gives
information on the wave function at the double and triple
coalescence points. Such a possible experimental access to the
triple coalescence point was not yet proposed in literature. In
the experiments \cite{5}-\cite{7}the projectiles which captured
two electrons were registered in coincidence with the photon. This
enables to distinguish the process among the other capture
processes in spite of its small cross section. Detection of the
process at small values of $\xi$ would be a bright manifestation
of the three particle singularity in a dynamical process.

Of course, Eq.~(\ref{4}) describes also the amplitude of double
photoionization with momenta of the outgoing electrons
$\bp_1=\bp_2$, $|\bp_i|\gg\eta$. It determines the double
differential cross section $\frac{d\sigma}{d\varepsilon d\tau}$ at
$\varepsilon_1=\varepsilon_2$, $\tau=(\bp_1\bp_2)/p_1p_2=1$,
which thus depends on the wave function behavior at
the triple coalescence point. However this region provides a
minor contribution to the differential cross section
$d\sigma/d\varepsilon$, which runs beyond the high energy limit.
The contribution to the double photoionization cross section is
still smaller. On the contrary, in the double electron
capture with single photon emission the triple coalescence
singularity determines the high energy limit of the cross section.

\subsection*{Acknowledgements}

We thank P. H. Mokler for useful discussions. The work was partially
supported by the DFG grant No.~436~RUS~113/822/. Two of us (E.~G.~D.
and A.~I.~M.) acknowledge hospitality during our visits to
Justus-Liebig-University of Giessen.


\end{document}